\documentclass[12pt]{article}
\usepackage{graphicx}

\begin{document}

\title{\bf Hamiltonian intermittency and L\'evy flights in the three-body problem}

\author{Ivan~I.~Shevchenko\/\thanks{E-mail:~iis@gao.spb.ru} \\
Pulkovo Observatory of the Russian Academy of Sciences \\
Pulkovskoje ave.~65, St.Petersburg 196140, Russia}
\date{}

\maketitle


\begin{center}
Abstract
\end{center}

\noindent
We consider statistics of the disruption and Lyapunov
times in an hierarchical restricted three-body problem. We show
that at the edge of disruption the orbital periods and the size of
the orbit of the escaping body exhibit L\'evy flights. Due to
them, the time decay of the survival probability is heavy-tailed
with the power-law index equal to $-2/3$, while the relation
between the Lyapunov and disruption times is quasilinear.
Applicability of these results in an ``hierarchical resonant
scattering'' setting for a three-body interaction is discussed.

\bigskip

\noindent Key words: three-body problem, Hamiltonian dynamics,
chaotic dynamics, Kepler map, Lyapunov exponents.

\newpage

\section{Introduction}

Notwithstanding a three-centennial progress in the studies of the
three-body problem, disruption of a three-body gravitational
system still remains an enigmatic dynamical process. The
statistics of the disruption times is of special interest. In
particular, it is interesting whether the computed Lyapunov
exponents of the motion are related in some way  to the disruption
times or, more generally, to the times of sudden changes in the
orbital behavior. (The inverse of the maximum Lyapunov exponent,
--- the Lyapunov time $T_\mathrm{L}$, gives the timescale of
exponential divergence of close trajectories in phase space; it
characterizes the level of predictability of the motion. The
disruption time $T_\mathrm{d}$ is the system lifetime as a bound
system.)

Much numerical-experimental work has been done on this subject in
the last two decades. A prominent relationship in the statistics
of sudden orbital changes in gravitational systems has been found
and confirmed. This relationship consists in the
close-to-quadratic character of the dependence of the time
$T_\mathrm{r}$ of a sudden orbital change on the Lyapunov time
$T_\mathrm{L}$. (The time of a sudden orbital change is
dynamically equivalent to the recurrence time for entering some
domain in the phase space of a Hamiltonian system, therefore it is
designated $T_\mathrm{r}$ here and in the following.) On the basis
of extensive numerical experiments in the dynamics of three-body
and more complicated gravitational systems (in particular, in
asteroidal dynamics), it was argued in \cite{SFL90,LFM92,MLF94}
that the time $T_\mathrm{r}$ could be statistically predicted by
means of computation of the maximum Lyapunov exponent. A simple
``universal'' statistical relationship was established to exist
between the computed time $T_\mathrm{r}$ and the computed Lyapunov
time $T_\mathrm{L}$ (the inverse of the computed maximum Lyapunov
exponent):
\begin{equation}
T_\mathrm{r} \propto T_\mathrm{L}^\beta \label{trtlfm}
\end{equation}

\noindent with $\beta = 1.7$--$1.8$ typically, though considerable
dispersion of the statistical data was usually present. The same
kind of relationship with $\beta \approx 1.9$ was found in
\cite{LD93} in simulations of the dynamics of the Kuiper belt of
asteroids. Similar dependences were found in \cite{FM97,S98a} for
the chaotic behavior of asteroids moving close to the mean motion
resonances 2/1 and 3/1 with Jupiter. Tsiganis et al.\ \cite{TVD05}
studied chaotic diffusion and effective stability of Jupiter
Trojans, and in a large chaotic region, surrounding the stability
zone in the motion of asteroids in the Trojan swarms, obtained a
statistical relationship between the escape and Lyapunov times,
similar to the ``universal'' close-to-quadratic one.

However, recent numerical-experimental statistical studies
\cite{MT07,UH08} of the disruption process in three-body systems
revealed a new --- quasilinear --- kind of relation between the
Lyapunov times and the disruption times. Mikkola and Tanikawa
\cite{MT07} explored correlation of the Lyapunov times
$T_\mathrm{L}$ and the disruption times $T_\mathrm{d}$ in the
equal-mass three-body problem. In extensive numerical experiments
they computed the disruption and Lyapunov times for the motion
with randomized initial conditions, and showed that the system
lifetimes, as a bound triple, and the Lyapunov times were
correlated, the ``$T_\mathrm{L}$~--- $T_\mathrm{d}$'' relationship
at long lifetimes being close to linear. Urminsky and Heggie
\cite{UH08} explored correlation between the Lyapunov and
disruption times in an hierarchical three-body problem in a
setting of the three-body problem different from that used in
\cite{MT07}, but obtained similar results. Namely, they considered
the Sitnikov problem \cite{S60} and, as an outcome of extensive
numerical experiments, obtained a two-part power-law relationship
with the second part (tail) close to linear one, similar to that
found in \cite{MT07} for the equal-mass three-body problem.

Though the results \cite{MT07,UH08} on the ``$T_\mathrm{L}$~---
$T_\mathrm{d}$'' relationship are similar, they diverge
significantly from the earlier results
\cite{SFL90,LFM92,MLF94,LD93,FM97,S98a,TVD05} obtained for the
statistics of sudden orbital changes.

In this paper we identify causes that lead to the quasilinear
``$T_\mathrm{L}$~--- $T_\mathrm{d}$'' relationship in the
three-body problem at the edge of disruption, as opposed to the
earlier found close-to-quadratic ``$T_\mathrm{L}$~---
$T_\mathrm{r}$'' dependence in resonant dynamics, and explain its
nature as an effect of Hamiltonian intermittency.

\vspace{5mm}

\section{Two kinds of Hamiltonian intermittency}
\label{hi}

The close-to-quadratic relationship~(\ref{trtlfm}) between the
recurrence and Lyapunov times in general Hamiltonian systems
with divided phase space (i.e., with regular and chaotic kinds of
motion both present) has a straightforward explanation if one
takes into account the so-called ``stickiness phenomenon''
\cite{S98a,S99}: the correlation arises because the trajectories
sticking to chaos border have large recurrence times (the times of
return to some domain of phase space), and, since they mimic
regular ones, they have also large computed (``local'') Lyapunov
times. The relationship between the recurrence time $T_\mathrm{r}$
and the local Lyapunov time $T_\mathrm{L}$ on each recurrence was
shown to be close to quadratic \cite{S98a}:
\begin{equation}
T_\mathrm{r} \propto {T_\mathrm{L}}^2, \label{trtl2}
\end{equation}

\noindent i.e., $\beta \approx 2$ in Eq.~(\ref{trtlfm}).
This relationship was derived in \cite{S98a} for the motion inside
the chaotic layer near the separatrices of nonlinear resonance in
the perturbed pendulum model, and this indeed covers a lot of
possible applications, including the chaotic motion of asteroids
near mean motion resonances. The separatrix map theory
\cite{C79,C90,LL92} was used for deriving relation~(\ref{trtl2}).

Generally speaking, the emergence of the algebraic
``$T_\mathrm{L}$~--- $T_\mathrm{r}$'' relationship is due to
Hamiltonian intermittency. The phenomenon of intermittency, i.e.,
chaotic behavior intermittently interchanged with close-to-regular
one, is well known in studies of dissipative dynamical
systems~\cite{LL92}. Much less it is discussed in connection with
Hamiltonian dynamics. There are two kinds of Hamiltonian
intermittency known. The first one \cite{ZSUC91} takes place when
the motion is ``stochastized'' at encounters with the separatrix,
while most of the time (far from the separatrix) the motion looks
regular. The separatrix in the simple model \cite{ZSUC91},
as well as in the
map models (\ref{km}, \ref{km1}) considered in this paper, is just
the singular line $y=0$ separating bound and unbound types of
motion. The first kind Hamiltonian intermittency is inherent to
adiabatic chaos; see~\cite{S08a,S08b} and references therein. The
second kind Hamiltonian intermittency \cite{S98b} takes place when
a fractal ``chaos--order'' boundary is present in phase space.
Sticking to the border results in long time segments of
close-to-regular behavior; they are interrupted by prominently
chaotic dynamical events, when the trajectory leaves the border
neighborhood \cite{SS96,SS97,S98a,S98b}. Thus the physical cause
for the Hamiltonian intermittency of the first kind consists in
close encounters of a trajectory with slowly pulsating separatrix,
while the physical cause for that of the second kind consists in
sticking of a trajectory to chaos border. Statistical properties
of these two kinds of phenomena are very different, as
demonstrated further in this paper. This is just the second kind
Hamiltonian intermittency that is responsible for the quadratic
``$T_\mathrm{L}$~--- $T_\mathrm{r}$'' relationship~(\ref{trtl2})
\cite{S98a}. In what follows we show that in certain circumstances
a quasilinear ``$T_\mathrm{L}$~--- $T_\mathrm{r}$'' relationship
can arise, and this is due to the first kind Hamiltonian
intermittency.

\section{General separatrix maps}
\label{gsm}

The nonlinear pendulum provides a model of nonlinear resonance
under definite conditions~\cite{C79,LL92}. The motion in the
vicinity of the separatrix of the nonlinear pendulum (nonlinear
resonance) is described by the separatrix map~\cite{C79,LL92}. We
write it in the form adopted in \cite{S98a}:
\begin{eqnarray}
     y_{i+1} &=& y_i + \sin x_i, \nonumber \\
     x_{i+1} &=& x_i + \lambda \ln \vert y_{i+1} \vert + c ,
\label{sm}
\end{eqnarray}

\noindent where $y$ denotes the normalized relative pendulum's
energy, $x$ is normalized time, the constants $\lambda$ and $c$
are parameters.

Consider a map similar to the separatrix map~(\ref{sm}), but with
a power-law phase increment instead of the logarithmic one:
\begin{eqnarray}
     y_{i+1} &=& y_i + \sin x_i, \nonumber \\
     x_{i+1} &=& x_i + \lambda \vert y_{i+1} \vert^{-\gamma} ,
\label{km}
\end{eqnarray}

\noindent or, in an equivalent form usually used,
\begin{eqnarray}
     w_{i+1} & = & w_i + W \sin \tau_i , \nonumber \\
     \tau_{i+1} & = & \tau_i + \kappa \vert w_{i+1} \vert^{-\gamma} .
     \label{km1}
\end{eqnarray}

\noindent Map (\ref{km1}) has two parameters, $W$ and $\kappa$,
instead of one parameter $\lambda$ in map~(\ref{km}); apart from
the $\gamma$ parameter. The two-parameter map~(\ref{km1}) is
reduced to the one-parameter map~(\ref{km}) with $\lambda = \kappa
W^{-\gamma}$ by means of straightforward substitution $w = W y$,
$\tau = x$.

A number of mechanical and physical models are described by maps
(\ref{km}) and (\ref{km1}) with rational values of $\gamma$. The
values of $\gamma = 1/4$ and $1/3$ correspond to the Markeev
maps~\cite{M95,M94} for the motion near separatrices of resonances
in two degenerate cases; $\gamma = 1/2$ gives the ``$\hat L$-map''
\cite{ZSUC91} for the motion of a non-relativistic particle in the
field of a wave packet, this value of $\gamma$ also gives a map
for the classical Morse oscillator driven by time-periodic force
\cite{A06}; $\gamma = 1$ gives the Fermi map \cite{ZC64,LL92} for
the Fermi acceleration mechanism for cosmic rays; $\gamma = 3/2$
gives the Kepler map~\cite{P86,CV86,VC88,PB88,A06} for a number of
physical and astronomical applications; $\gamma = 2$ gives the
``ultrarelativistic map'' \cite{ZSUC91} for the motion of a
relativistic particle in the field of a wave packet. All these
maps describe dynamical behavior in the vicinities of separatrices
of corresponding models. In the case of the Kepler map the
separatrix (the line $y=0$) separates the bound and unbound states
of motion.

The Kepler map (Eqs.~(\ref{km}) or (\ref{km1}) with $\gamma =
3/2$) was derived and analyzed in \cite{P86,CV86,VC88,PB88} in
order to describe the chaotic motion of the Halley comet and,
generally, the motion of comets in highly eccentric orbits. The
motion model for the Kepler map consists in the assumption that
the main perturbing effect of planets (Jupiter first of all) is
concentrated when the comet is close to the perihelion of its
orbit. The $y$ variable has the meaning of the normalized orbital
energy of the comet, while $x$ is the normalized time. One
iteration of the map corresponds to one orbital revolution of the
comet. This means that the map time unit, corresponding to one
iteration, is not constant. The increment of real time per
iteration is $\Delta x_i = x_i - x_{i-1}$.

The Kepler map is known to describe dynamics in several different
settings of an hierarchical restricted three-body problem: in the
external restricted planar \cite{P86,CV86,PB88} and strongly
non-planar \cite{E90} problems in cometary dynamics; also in the
abstract Sitnikov problem, where the tertiary moves along the {\it
perpendicular} to the orbital plane of the main binary. The
validity of the map in the last case follows from the work by
Urminsky and Heggie
\cite{UH08}, who considered a variant of the Sitnikov problem and
derived a map (see Eqs.~(11) in \cite{UH08}), describing the
dynamics in this problem. This map is straightforwardly reducible
to the Kepler map (\ref{km}); in particular, using formulas given
in \cite{UH08} for the parameters of their map, it is easy to
derive a formula for the parameter $\lambda$ in Eqs.~(\ref{km}):
\begin{equation}
\lambda = 2 \pi (2.029 e)^{-3/2} \approx 2.17 e^{-3/2} ,
\label{kappa}
\end{equation}

\noindent where $e$ is the eccentricity of the central binary. So,
for the eccentricity $e$ equal to $0.1$ and $0.6$ (chosen as
representative in \cite{UH08}) one has $\lambda \approx 70$ and
$5$, respectively. We choose these two values of $\lambda$ for
illustrative numerical experiments in what follows.

\section{L\'evy flights at the edge of escape: \\ the
distribution}
\label{lfeed}

The L\'evy flights, i.e., the increments (in a kind of random
walk) that have a heavy-tailed distribution, is a well-studied
subject with various applications. In Hamiltonian dynamics, they
were thoroughly considered in connection with what we call the
second kind Hamiltonian intermittency; see, e.g, \cite{DKU02} and
references therein. In celestial mechanics, L\'evy random walks
arising due to close encounters of bodies were considered in the
orbital energy evolution of comets \cite{ZSZ02}.

Usually, L\'evy flights are considered in random walks with steps
possible in at least two (forward and back) directions; but the
L\'evy distributions considered below are one-sided: the
increments (the orbital periods of the escaping body and the
recurrence times) are positive. In a general statistical setting,
one-sided L\'evy flights were considered in \cite{KKM07,K07},
where exact results for the first passage time and leapover
statistics were obtained.

Both kinds of L\'evy flights explored below in the framework of
the three-body problem are {\it not} due to encounters of bodies.
One of these two kinds is due to encounters of a trajectory with
the separatrix in phase space; hereafter we call such flights the
``L\'evy flights of the first kind'', or LF1, since this
phenomenon corresponds to the Hamiltonian intermittency of the
first kind. L\'evy flights of another kind, which we call the
``L\'evy flights of the second kind'', or LF2, are due to sticking
of a trajectory to chaos border. In other words, these flights
arise due to the second kind Hamiltonian intermittency. They
effect the duration of long Poincar\'e recurrences, but not
(practically) the orbital periods.

In the orbital behavior of the tertiary in the planar restricted
three-body problem which we consider in what follows, the LF1
appear as sudden jumps in the orbital size and period, while LF2
appear as long sequences of orbital revolutions with almost
constant orbital size and eccentricity.

LF1 still take place when encounters of bodies are impossible.
E.g., in the cometary dynamics model, described by the Kepler map,
\cite{P86,CV86} the perihelion distance of the comet can be
greater than the semi-major axis of the orbit of the secondary by
any amount, but encounters with the separatrix with or without
crossing it can still take place (if the cometary orbit is
chaotic) and, as we shall see below, shape the statistics of
recurrences.

\begin{figure}[h!]
\centering
\includegraphics[width=0.75\textwidth]{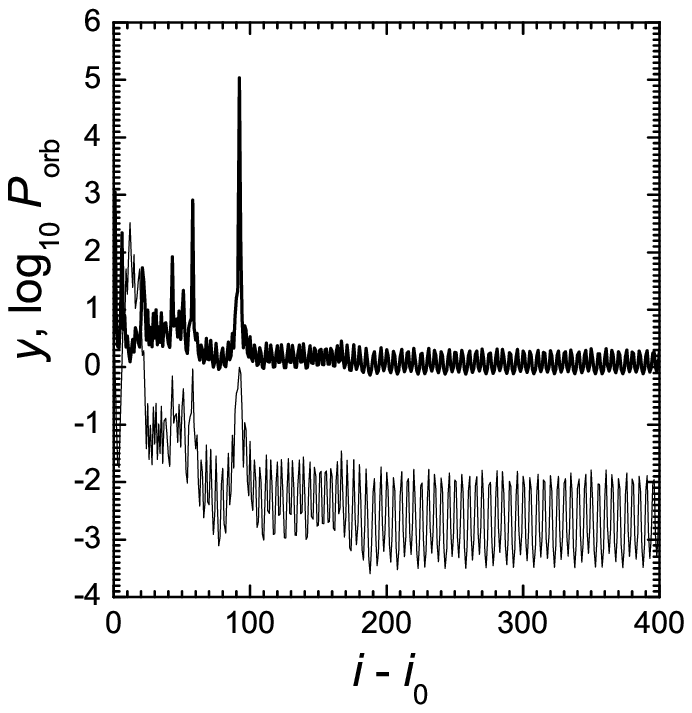}
\caption{A fragment of a chaotic trajectory of the Kepler map,
demonstrating L\'evy flights of the first kind (the prominent
peaks of $\log_{10} P_{\rm orb}$ on the left) and a L\'evy flight of
the second kind (the oscillatory ``plateau'' of $y$ and
$\log_{10} P_{\rm orb}$ on the right). $y$ is shown with a thin line,
and $\log_{10} P_{\rm orb}$ with a bold one; $\lambda = 5$; the value
of $i_0$ is some big number chosen to exhibit this part of
trajectory.}
\label{lf}
\end{figure}

In the dynamics of map~(\ref{km}), the LF1 and LF2 coexist (if
$\lambda$ is large enough, this condition is considered below). In
Fig.~\ref{lf}, a fragment of a chaotic trajectory of
map~(\ref{km}) with $\gamma = 3/2$ is shown, demonstrating LF1
(narrow peaks in the variation of the orbital period $P_{\rm orb}$
of the tertiary, on the left of the graph) and an LF2 (the
oscillatory low ``plateau'' in the variation of $y$ and $P_{\rm
orb}$, on the right). When $y$ hits close to zero, a jump in the
orbital period is observed. When the trajectory sticks to chaos
border, both the energy and the orbital period oscillate near some
low constant value.

The stickiness effect determines the character of the distribution
of Poincar\'e recurrences on large timescales: it is algebraic
\cite{CS81,CS84}. The algebraic decay in the recurrence statistics
in Hamiltonian systems with divided phase space was considered, in
particular, in \cite{CS81,CS84,C90,C99,CK08}, starting with a
pioneering work by Chirikov and Shepelyansky \cite{CS81}. Chirikov
\cite{C90}, using his resonant theory of critical phenomena in
Hamiltonian dynamics, justified a value of $3/2$ for the critical
exponent $\alpha$ in the integral distribution
\begin{equation}
F \propto T_\mathrm{r}^{-\alpha} \label{fta}
\end{equation}

\noindent of recurrences. The integral distribution
$F(T_\mathrm{r})$ is defined here as the relative share of the
recurrences with the duration greater than $T_\mathrm{r}$ in the
whole sample. In a recent paper \cite{CK08} the algebraic decay of
Poincar\'e recurrences was explored statistically on the basis of
large computational data on behavior of various Hamiltonian
systems. These numerical experiments showed system-dependent
power-law exponents, but the mean ``universal'' exponent turned
out to be well-defined and equal to $1.57 \pm 0.03$, somewhat
differing from the standard $3/2$ value. Venegeroles
\cite{V09} reports a value, equal to $1.54 \pm 0.07$, resulting
from averaging independent results of numerical studies of a
number of various Hamiltonian systems; see Table~1 and references
in \cite{V09}.

In celestial mechanics, the algebraic decay was observed in
numerical experiments on asteroidal dynamics \cite{SS96,SS97}.
These experiments were performed in the framework of the
restricted three-body problem Sun--Jupiter--asteroid. It was
shown that the tail of the distribution of duration $T_\mathrm{r}$
of intervals between jumps of the orbital eccentricity of
asteroids in the 3/1 mean motion resonance with Jupiter is
algebraic:
\begin{equation}
F \propto T_\mathrm{r}^{-\alpha}
\end{equation}

\noindent with $\alpha \approx 1.5$--$1.7$ typically. This
was interpreted in \cite{SS97} as an effect of sticking of the
chaotic orbits to chaos border in the divided phase space. In
other words, this is an effect of the second kind Hamiltonian
intermittency.

On the other hand, Dones et al.\ \cite{DLD96} studied the escape
times in the highly-eccentric chaotic cometary dynamics in the
Solar system (the perturbations due to the four giant planets were
taken into account) and reported on the algebraic tails of the
integral distributions with the power-law index equal to $0.8 \pm
0.2$. This behavior has not been theoretically interpreted up to
now; we shall see that it can be straightforwardly interpreted as
an effect of the first kind Hamiltonian intermittency.

The decay with $\alpha \approx 1.5$ is expected when LF2 dominate
over LF1 in the long recurrences. A different kind of the decay
law was observed in the behavior of the Kepler map by Borgonovi et
al.\ \cite{BGS88}. They analyzed long-time decay properties of a
Kepler map describing a one-dimensional model of hydrogen atom in
a microwave field, and by means of
rigorous deduction found a $T_\mathrm{d}^{-2/3}$ law for the
first approximation for the time decay of the survival probability
in the case of the escape times measured in real (constant) time
units. This law was confirmed by them in computed statistics; see
Fig.~1 in \cite{BGS88}. On the opposite, when the escape times
were measured in map (fictitious) time units, the usual
$T_\mathrm{d}^{-3/2}$ law was observed; see Fig.~2 in
\cite{BGS88}.

Apart from the rigorous treatment \cite{BGS88}, several heuristic
deductions of the $T_\mathrm{d}^{-2/3}$ law are available in
relevant problems \cite{SB01,H93,MT99}. Hut \cite{H93} derived an
heuristic $T_\mathrm{d}^{-2/3}$ law as a lower bound for the time
decay of the survival probability in a general ``hierarchical resonant
scattering'' \cite{H75,H93} setting for a three-body interaction
(where the masses of ``stars'' are arbitrary).
Malyshkin and Tremaine \cite{MT99} derived the
$T_\mathrm{d}^{-2/3}$ law for the time decay of the survival
probability for cometary ensembles in the Solar system. Schlagheck
and Buchleitner \cite{SB01} derived the $T_\mathrm{d}^{-2/3}$ law
for the time decay of the survival probability in an autoionizing
configuration of chaotic helium.

In all these approaches, the $T_\mathrm{d}^{-2/3}$ law was derived
without using any Kepler map. Two basic assumptions were always
made explicitly or implicitly, (i)~that the distribution of
ejection energies is flat or smooth in the neighborhood of the
energy threshold $E=0$, (ii)~that the asymptotic decay of the survival
probability is the same as the tail of the distribution of the
orbital periods of the escaping body. While the first assumption
is reasonable (in view of the complete ergodicity of the motion
near the threshold, see analysis given below for the Kepler map),
the second one is solely hypothetical. E.g., the argumentation
presented in \cite{MT99} is that the escaping comets ``remain
bound until their second perihelion passage, after which they will
normally be ejected within a relatively short time''. However,
given that all results \cite{SB01,H93,MT99} coincide with the
result of the rigorous treatment in \cite{BGS88}, we can infer
that the second assumption seems to be also valid.

Let us, using the same two assumptions, find a law for the
asymptotic distribution of the phase increments per iteration
(which are the orbital periods of the escaping body in case of
$\gamma = 3/2$) in the general case of arbitrary $\gamma$ in
map~(\ref{km}). The phase increment is
\begin{equation}
P = \Delta x_i = x_i - x_{i-1} = \lambda \vert y_i \vert^{-\gamma}
\label{tg}
\end{equation}

\noindent in real time units. In case of $\gamma = 3/2$ the real
time unit is equal to the orbital period of the central binary,
divided by $2 \pi$.

So, $\vert y_i \vert = \lambda^{1/\gamma} P^{-1/\gamma}$.
The motion in the close vicinity of the separatrix is locally
ergodic, i.e., regular islands are absent. This follows from the
fact that in the local (in the energy $y$) approximation of
map~(\ref{km}) by the standard map the stochasticity parameter of
the standard map tends to infinity when $y$ tends to zero. The
ergodicity implies that, for the $y$ variable close to zero, the
distribution function of $y$ is flat: $f(y) =$ const. One has
\begin{equation}
\mbox{\rm const} \ \mbox{\rm d} \vert y \vert \propto
P^{-\frac{1}{\gamma} - 1} \mbox{\rm d} P ,
\label{dfte}
\end{equation}

\noindent and the differential distribution function of $P$ is
\begin{equation}
f(P) \propto P^{-\frac{1}{\gamma} - 1} .
\label{dft}
\end{equation}

\noindent The integral distribution is $F(P) \propto
P^{-\frac{1}{\gamma}}$.

For the Kepler map, $\gamma = 3/2$ and $P = P_{\rm orb}$
(the orbital period); so, the differential
distribution is $f(P_{\rm orb}) \propto P_{\rm orb}^{- 5/3}$ and
the integral distribution is $F(P_{\rm orb}) \propto P_{\rm
orb}^{-2/3}$. From Kepler's third law one has $f(a) \propto
a^{-2}$ for the differential distribution of the semi-major axis
of the orbit; i.e., the distribution of the orbital size is also
heavy-tailed, and the L\'evy flights are demonstrated in the
process of disruption both in tertiary's orbital period and size.

By assumption~(ii), advocated above, law~(\ref{dft}) coincides
with the asymptotics of the time decay of the survival
probability. Alternatively, the same law follows from repeating
the analytical treatment \cite{BGS88} for the case of arbitrary
$\gamma$ instead of the exponent $3/2$ in the Kepler map: if one
repeats the rigorous deduction \cite{BGS88} (performed in
\cite{BGS88} for the Kepler map, $\gamma = 3/2$) in the general
case of arbitrary $\gamma$, one finds the distribution
$F(T_\mathrm{r}) \propto T_\mathrm{r}^{-\frac{1}{\gamma}}$ for the
recurrence times.
Note that the basic assumption in this deduction is that the
phases are randomized after each kick.

Borgonovi et al.\ \cite{BGS88} explained the LF1 dominance over
LF2 (in our terms) in real time statistics in the dynamics of the
Kepler map, using argumentation based on the infinite measure of
extended phase space near the separatrix. If one uses this
argumentation in the general case of map~(\ref{km}) with arbitrary
value of $\gamma > 0$, it follows that LF1 should dominate
for all $\gamma \geq 1$ at least. However, while LF1 dominate indeed
at $\gamma \geq 1$, the transition to this domination occurs at a
value less than 1. Let us estimate this critical value. At $\gamma
< \gamma_\mathrm{crit}$, when LF2 dominate, the slope index is
critical (in a different sense, related to the ``critical''
structure at chaos border, see discussion above): $\alpha =
\alpha_\mathrm{crit} \approx 3/2$. At $\gamma >
\gamma_\mathrm{crit}$, when LF1 dominate, the slope index $\alpha
= 1/\gamma$. These two curves $\alpha(\gamma)$ intersect at
$\gamma = \gamma_\mathrm{crit} = 1 / \alpha_\mathrm{crit} \approx
2/3$.

What is the physical reason for the switch to take place at this
point? For the recurrences forming LF2 with duration greater than
$T$, the ``total sojourn time'' $\sim T F(T)$ \cite{C99}, i.e.,
$\sim T^{-\alpha_\mathrm{crit} + 1}$. Analogously, in the LF1
case, the total sojourn time is $\sim T F(T) \propto
T^{-\frac{1}{\gamma} + 1}$. LF1 asymptotically dominate, if the
second sojourn time is greater than the first one:
$T^{-\frac{1}{\gamma} + 1} > T^{-\alpha_\mathrm{crit} + 1}$. Hence
the condition for the LF1 domination is $\gamma >
1/\alpha_\mathrm{crit}$. This is exactly what we have just derived
for the point of intersection of the two curves $\alpha(\gamma)$.

So, a critical non-trivial value $\gamma_\mathrm{crit}$ of the
$\gamma$ parameter exists (if $\lambda$ is large enough, see
below), such as the maps with $\gamma > \gamma_\mathrm{crit}$ have
LF1 dominating over LF2 in real time statistics, whereas at
$\gamma < \gamma_\mathrm{crit}$ LF2 dominate in both the real time
and map time statistics.

The quantity $\gamma_\mathrm{crit}$ equals $2/3$ in the case of
the standard value $\alpha = 3/2$; and $\gamma_\mathrm{crit}
\approx 0.637$ in the case of $\alpha = 1.57$ computed in
\cite{CK08}. In particular, LF1 dominate over LF2 in real time
statistics in the long Poincar\'e recurrences in the dynamics of
the Fermi and Kepler maps, while for the ordinary separatrix maps,
Markeev maps and $\hat L$-maps, if $\lambda$ is large enough, the
tails of the recurrence distributions are LF2-dominated and their
slopes do not depend on the choice of units (map units or real
time units) in which the lengths of recurrences are measured.

\begin{figure}[h!]
\centering
\includegraphics[width=0.75\textwidth]{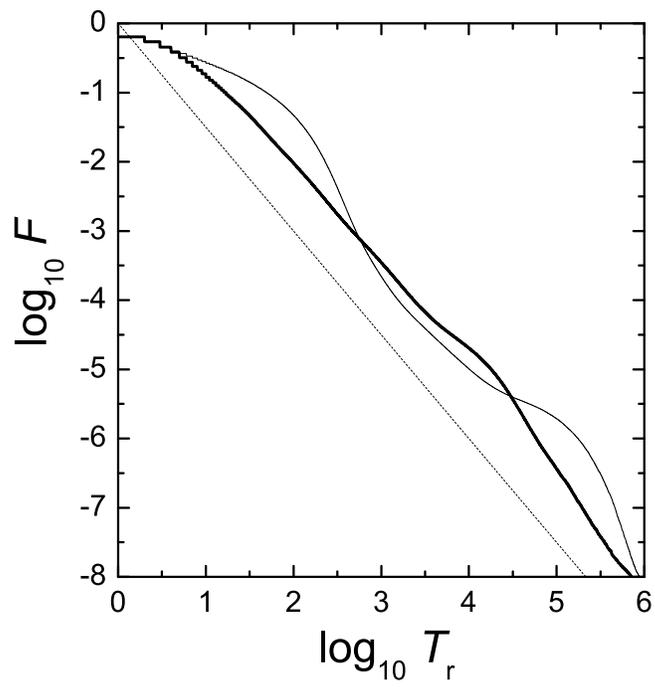}
\caption{Integral distributions of the recurrence times measured
in map time units (iterations) for the Kepler map; $\lambda = 5$
{\it (bold line)} and $\lambda = 70$ {\it (thin line)}. The
straight dotted line shows the $T_\mathrm{r}^{-3/2}$ law.}
\label{dism}
\end{figure}

\begin{figure}[h!]
\centering
\includegraphics[width=0.75\textwidth]{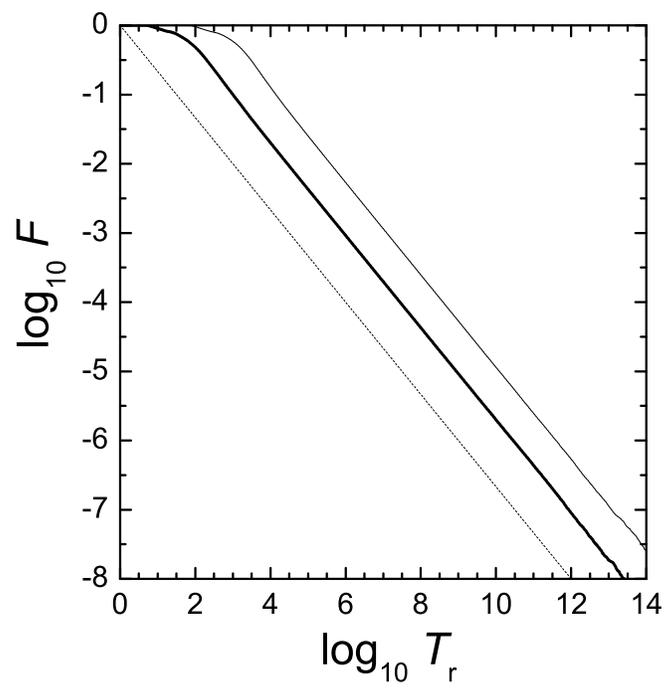}
\caption{The same as in Fig.~\protect{\ref{dism}}, but the
recurrence times are measured in real time units. The bold line
corresponds to $\lambda = 5$, and the thin one to $\lambda = 70$.
The straight dotted line shows the $T_\mathrm{r}^{-2/3}$ law.}
\label{disc}
\end{figure}

To illustrate the difference between distributions of recurrences
measured in map and real time units, we present here examples of
distributions obtained by iterating map (\ref{km}) with $\gamma =
3/2$ (the Kepler map). In Fig.~\ref{dism}, the computed integral
distributions $F(T_\mathrm{r})$ of the recurrence times
$T_\mathrm{r}$ measured in map time units (iterations), are shown
for $\lambda = 5$ {\it (bold line)} and $\lambda = 70$ {\it (thin
line)}. The number of iterations $n_{\rm it} = 10^{11}$ in both
cases. The recurrences are counted at the line $y = 0$. The
quantity $F(T_\mathrm{r})$ is the fraction of recurrences longer
than $T_\mathrm{r}$. The tail of the distribution in
Fig.~\ref{dism} follows approximately the power law with the slope
index $\alpha \approx 1.5$, as expected. It is similar, e.g., to
the tails of distributions presented in Fig.~4 in \cite{SS97} for
the intervals between eccentricity bursts of chaotic asteroidal
trajectories in the 3/1 Jovian resonance, where the same critical
dynamical mechanism (sticking to chaos border) is present. In
Fig.~\ref{disc}, the same distributions as in Fig.~\ref{dism} are
shown, but the recurrence times are measured in real time units.
The slope index for the tails is evidently equal to $2/3$, as
expected in this case. So, when real time units are used, the
tails of the distributions demonstrate a behavior that is
different in slope and regularity, compared to the case of using
the map time units. The reason is that LF1 dominate in the
distribution tail in the first case, while LF2 dominate in the
second case.

As we have already mentioned above, the LF1 and LF2 coexist in the
dynamics of map~(\ref{km}), if $\lambda$ is large enough. This
condition provides the existence of the global fractal chaos
border and, consequently, the prominent sticking phenomenon. If
$\lambda$ is small enough, map~(\ref{km}) can be reduced to a
differential equation describing regular trajectories, at all $y$
far from the separatrix $y=0$. This can be done analogously to the
case of the ordinary separatrix map~(\ref{sm}); the procedure is
described in \cite{S08a}. Thus the motion far from the separatrix
is locally regular, no global fractal chaos border exists; only
LF1 are possible. This is the realm of ``adiabatic chaos''.

What is the boundary value of $\lambda$, separating the cases with
and without global fractal chaos border? It can be estimated from
the form of the $\lambda$ dependence of the maximum Lyapunov
exponent $L$ of map~(\ref{km}). The $L$ value increases with
$\lambda$, while $\lambda$ is small, but then saturates at some
constant value (see \cite[fig.~3]{S07} for the Kepler map case
$\gamma = 3/2$). The saturation takes place when the role of the
global fractal chaos border becomes important in the dynamics. As
follows from \cite[fig.~3]{S07}, the boundary value of $\lambda$
for this map is $\approx 2$--3. In the case of the ordinary
separatrix map, the transition value of $\lambda$ is $\approx
0.5$--1 (see \cite[fig.~3]{S04}, also \cite[fig.~1]{S08a}). By
means of constructing the dependences $L(\lambda)$ for arbitrary
$0 < \gamma < 2$, it can be shown that the boundary value of
$\lambda$ does not change much with $\gamma$ and, by the order of
magnitude, is $\sim 1$. Thus, if $\lambda \gg 1$, we expect
coexistence of LF1 and LF2 in the dynamics of map~(\ref{km}),
while if $\lambda \ll 1$, only LF1 are possible.

\begin{figure}[h!]
\begin{tabular}{c}
{\bf a)} \includegraphics[width=0.6\textwidth]{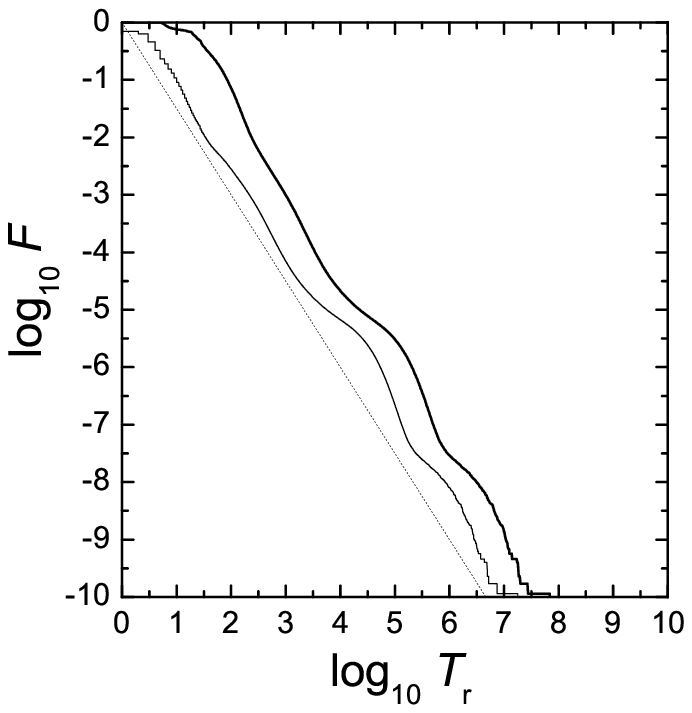} \\
{\bf b)} \includegraphics[width=0.6\textwidth]{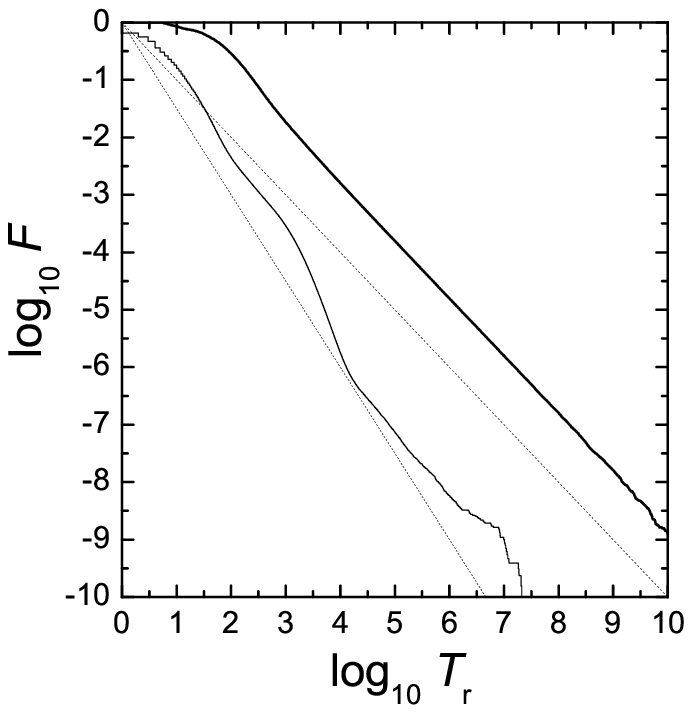} \\
\end{tabular}
\caption{Integral distributions of the recurrence times measured
in real time {\it (bold lines)} and map time {\it (thin lines)}
units for the $\hat L$-map (a) and for the Fermi map (b); $\lambda
= 5$; $n_{\rm it} = 10^{11}$. The straight dotted lines show the
$T_\mathrm{r}^{-3/2}$ and $T_\mathrm{r}^{-1}$ laws.}
\label{L_F_maps}
\end{figure}

Taking this into account (i.e., setting $\lambda$ to be large
enough), one can exploit the phenomenon of coexistence of LF1 and
LF2 to estimate the value of the critical exponent $\alpha$.
Namely, one can compute the distribution functions (in real time)
of the Poincar\'e recurrences for map~(\ref{km}) for a set of
values of $\gamma$ with some step, and, on increasing $\gamma$,
fix its transition value $\gamma_\mathrm{crit}$, when the tail of
the integral distribution function starts to take the form
characteristic for the LF1 statistics, i.e., the algebraic decay
with the power-law index $\alpha$ equal to $1/\gamma$. Then
$\alpha = 1 / \gamma_\mathrm{crit}$. Thus the procedure gives a
new method for estimating the critical exponent $\alpha$ in
Eq.~(\ref{fta}). An example is given by the graphs in
Fig.~\ref{L_F_maps}. One can see that the realization of our
method with only two values of $\gamma$, namely, $\gamma=1/2$
(corresponding to the $\hat L$-map) and $\gamma=1$ (corresponding
to the Fermi map), clearly shows that $1 < \alpha < 2$.

Now let us investigate this effect in more detail by building
the dependence $\alpha(\gamma)$ directly, i.e., by finding
$\alpha$ numerically on a grid of values of $\gamma$. When LF2
dominate, this dependence is expected to be subject to
deformations arising due to marginal resonances at chaos border:
at some intervals of $\gamma$, the border of the chaotic layer
and, consequently, the recurrence statistics are perturbed due to
emergence of the marginal resonances; on the marginal resonances,
see \cite{S98b}. In order to reduce maximally these perturbations,
let us introduce a constant shift $c$ in $x$ in map~(\ref{km}), in
the following way:
\begin{eqnarray}
     y_{i+1} &=& y_i + \sin x_i, \nonumber \\
     x_{i+1} &=& x_i + \lambda \vert y_{i+1} \vert^{-\gamma} + c.
\label{kms}
\end{eqnarray}

\noindent The shift $c$ is a new parameter, analogous to the
parameter $c$ in map~(\ref{sm}). At each value of $\gamma$ we
adjust the value of $c$ in such a way that the winding number of
the last undestroyed global KAM curve at the border of the chaotic
layer is approximately equal to the ``golden mean'' $(5^{1/2} -
1)/2$; in other words, it is maximally far from the main resonances.

By means of linearization of map~(\ref{km}) or (\ref{kms}) in $y$
it is straightforward to see that the value of $y$ corresponding
to the critical value of the stochasticity parameter $K = K_{\rm
G} \approx 0.971635406$ \cite{C79,LL92,M92} of the approximating
standard map is $y_{\rm b} = (\gamma \lambda / K_{\rm G})^{1 \over
{\gamma + 1}}$. This value corresponds to the border of the
chaotic layer. The regularizing constant shift $c$ is found by
averaging the phase increment at the border of the chaotic layer;
the resulting formula is
\begin{equation}
c_\mathrm{reg} = \pi (5^{1/2} - 1) - \lambda (\gamma
\lambda/K_\mathrm{G})^{-\gamma/(\gamma + 1)} . \label{creg}
\end{equation}

Since the form of the relationship $\alpha(\gamma)$ is expected to be
independent from $\lambda$, when $\lambda$ is large enough, one can
take any large value $\lambda \gg 1$; we choose $\lambda=10$. We
build the graph on the interval $0 < \gamma \leq 2$ with the
resolution $\Delta \gamma = 0.01$, i.e., we make 200 measurements
of $\alpha$. At each value of $\gamma$ on the grid, the integral
distribution built in logarithmic coordinates is linearly fitted
on the interval $3 < \log_{10} T_\mathrm{r} < 6$; $T_\mathrm{r}$
is measured in real time units. The left border of the interval is
chosen to be much greater than the point of transition from the
initial behavior (which can be exponential or inverse square-root,
see \cite{C99}) to the asymptotic one. A direct inspection of the
constructed distributions shows that this transition occurs at
$\log_{10} T_\mathrm{r} < 2$ for all points on the $\gamma$ grid.
The right border of the $T_\mathrm{r}$ interval is defined to be
much less than the values at which the distributions exhibit a
drop due to poor statistics at the very tails. At each point on
the $\gamma$ grid, the number of iterations of the map is
$n_\mathrm{it} = 10^{10}$.

\begin{figure}[h!]
\centering
\includegraphics[width=0.75\textwidth]{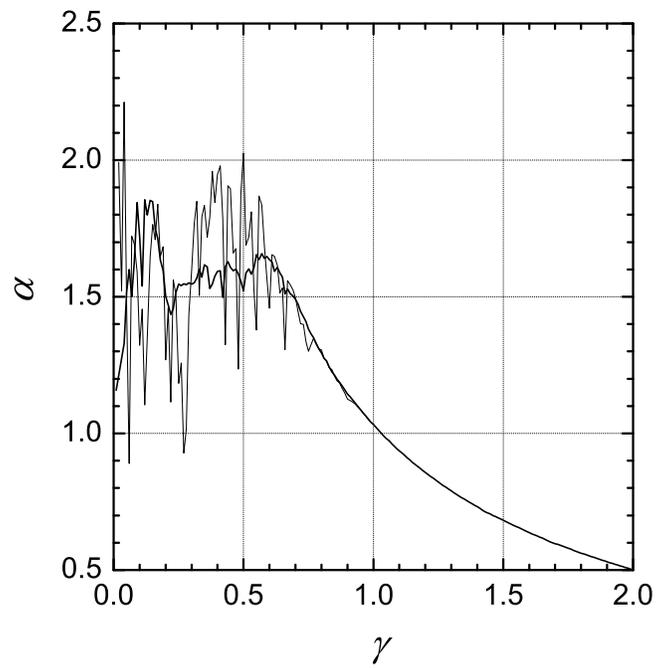}
\caption{The dependence of the computed power-law index $\alpha$
on the $\gamma$ parameter, for $c=c_\mathrm{reg}$ (bold line) and
$c=0$ (thin line); $\lambda=10$.}
\label{ag}
\end{figure}

In Fig.~\ref{ag}, the computed dependence $\alpha(\gamma)$ is
shown for two cases, $c=0$ and $c=c_\mathrm{reg}$. The {\it bold line}
corresponds to the map with $c=c_\mathrm{reg}$, whereas the {\it thin
line} corresponds to the map with $c=0$. One can see that in the
first case the irregular perturbations of the dependence are
indeed suppressed seriously, though they are not at all completely
eliminated. In the both cases, the transition to the LF1
domination at $\gamma \approx 0.7$ is evident, in agreement with
our prediction. However, it is also evident that the numerical
construction of the $\alpha(\gamma)$ dependence does not provide a
high-precision tool for determining the $\alpha_\mathrm{crit}$
value, because the perturbations of the theoretically ``flat''
behavior at $\gamma < \gamma_\mathrm{crit}$ are large, and, due to
them, the transition point cannot be located precisely. What is
more, the suppression of the border perturbations by adjustment of
the $c$ parameter might introduce systematic errors
(yet unexplored theoretically) in the estimated value of
$\alpha_\mathrm{crit}$.

\begin{figure}[h!]
\centering
\includegraphics[width=0.75\textwidth]{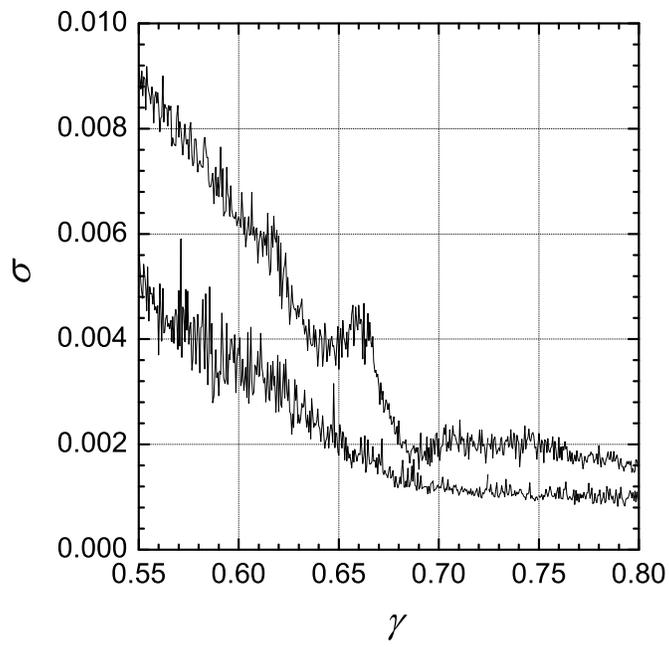}
\caption{The $\gamma$ dependence of the standard deviation
$\sigma$ for the computed power-law index $\alpha$; $\lambda=5$
(lower line) and $\lambda=10$ (upper line); $c=c_\mathrm{reg}$.}
\label{sag}
\end{figure}

Another way seems to be much more promising here. To locate the
transition point, one can exploit the property of fluctuations of
the recurrence time distribution in the case of the second kind
Hamiltonian intermittency, as opposed to the perfectly unperturbed
behavior in the case of its first kind (see Figs.~\ref{L_F_maps}).
In our fits of the $T_\mathrm{r}$ distributions, the
$\alpha$ values are determined with their standard errors, i.e.,
at each point of $\gamma$ one obtains $\alpha \pm \sigma$ as an
estimate for the power-law index. In Fig.~\ref{sag},
high-resolution $\gamma$ dependences of the standard deviation
$\sigma$ for the numerically determined power-law index $\alpha$
are shown for $\lambda=5$ {\it (lower line)} and $\lambda=10$
{\it (upper line)}; $c=c_\mathrm{reg}$.
The interval of $\gamma$ is taken
to be in the neighborhood of the expected switch value; namely,
$0.55 \leq \gamma \leq 0.80$. The resolution of the plot in
$\gamma$ is $\Delta \gamma = 0.0005$. All other details of
the numerical procedure are the same as adopted for
construction of Fig.~\ref{ag}.

In the constructed dependences in Fig.~\ref{sag}, a transition from
a linear decline from large values of $\sigma$ to a low-level
horizontal plateau takes place
in a narrow interval $\approx 0.67 < \gamma < \approx 0.68$
(corresponding to $\alpha_\mathrm{crit} \approx 1.5$), in agreement
with our prediction for the critical value of $\gamma$.
Concluding, construction and analysis of the dependence
$\sigma(\gamma)$ does indeed provide a perspective numerical tool
for determining the $\alpha_\mathrm{crit}$ value.

\section{L\'evy flights at the edge of escape: \\
the ``$T_\mathrm{L}$~--- $T_\mathrm{r}$'' relation} \label{lfeett}

As discussed in the Introduction, the power-law character of the
dependence of the times of sudden orbital changes on the Lyapunov
times~\cite{SFL90,LFM92,MLF94,LD93,FM97,S98a,TVD05} was explained
in~\cite{S98a,S99} as a phenomenon of critical dynamics, i.e., an
effect of motion near chaos border. This is an effect of the
second kind Hamiltonian intermittency.

Here we show that the first kind Hamiltonian intermittency in the
three-body problem at the edge of disruption leads to a
quasilinear ``$T_\mathrm{L}$~--- $T_\mathrm{r}$'' relation, when
the times are measured in real time units. As follows from
Eqs.~(\ref{km}), the length of a Poincar\'e recurrence in real
time units is
\begin{equation}
T_\mathrm{r}^{{\rm (ru)}} = \sum_{i=1}^n \Delta x_i = \lambda
\sum_{i=1}^n \vert y_i \vert^{-3/2}, \label{ttqc}
\end{equation}

\noindent where $n = T_\mathrm{r}$ is the duration of the
Poincar\'e recurrence in map time units (iterations); the time
increment $\Delta x_i = x_i - x_{i-1}$, and the iteration $i=1$ is
set for the start of the recurrence.

Let us consider the finite-time maximum Lyapunov exponent (of an
original dynamical system) calculated for a recurrence in the case
when real time units are used. To obtain it, one should divide the
calculated finite-time maximum Lyapunov exponent $L$ of the map by
the average (on the recurrence) length $\langle P_{\rm map}^{{\rm
(ru)}} \rangle = T_\mathrm{r}^{{\rm (ru)}}/T_\mathrm{r}$ of the
map iterations in real time (see, e.g., \cite{LL92}). Hereafter we
denote $\langle P_{\rm map}^{{\rm (ru)}} \rangle$ by $q$:
\begin{equation}
q = \frac{T_\mathrm{r}^{\rm (ru)}} {T_\mathrm{r}} = \frac{1}{n}
\sum_{i=1}^n \Delta x_i = \frac{\lambda}{n} \sum_{i=1}^n \vert y_i
\vert^{-3/2} . \label{q}
\end{equation}

\noindent The maximum Lyapunov exponent referred to real time
units is
\begin{equation}
L^{{\rm (ru)}} = \frac{L}{q} . \label{lcu}
\end{equation}

\noindent Equivalently, for the Lyapunov time one has
\begin{equation}
T_\mathrm{L}^{{\rm (ru)}} = q T_\mathrm{L} . \label{tlcu}
\end{equation}

\noindent On the other hand,
\begin{equation}
T_\mathrm{r}^{{\rm (ru)}} = q T_\mathrm{r} . \label{trcu}
\end{equation}

\noindent The quantity $q = T_\mathrm{r}^{\rm (ru)}/T_\mathrm{r}$
is the ratio of two random variables, distributed as power laws,
as shown above. If the $y$ value hits close to the separatrix,
then, according to (\ref{ttqc}), there is a jump in
$T_\mathrm{r}^{\rm (ru)}$, but there is no jump in $T_\mathrm{r}$.
If the magnitude of such jumps is much greater than the total
range of the original ``$T_\mathrm{L}$~--- $T_\mathrm{r}$'' graph
in map time units (this is the case when LF1 dominate over LF2),
the graphical relationship in real time units will be spread, due
to the jumps, in the direction $T_\mathrm{L} = T_\mathrm{r}$. So,
when $T_\mathrm{L}$ are $T_\mathrm{r}$ are expressed in real time
units and LF1 dominate over LF2, the ``$T_\mathrm{L}$~---
$T_\mathrm{r}$'' relationship becomes quasilinear.

This inference is valid when LF1 dominate. For the ordinary
separatrix map (\ref{sm}) and general maps (\ref{km}) with $\gamma
< \gamma_\mathrm{crit}$, where singularity is weaker, LF2 dominate
if $\lambda$ is large enough, and the generic relationship is not
spread in the $T_\mathrm{L} = T_\mathrm{r}$ direction. This is the
reason why the close-to-quadratic relationship
\cite{SFL90,LFM92,MLF94,LD93,FM97,S98a,TVD05}, and not the
quasilinear one, is present in the dynamics of minor Solar system
bodies near resonances, where the ordinary separatrix map, but not
the Kepler map, is relevant to the dynamics of interacting
nonlinear resonances.

\begin{figure}[h!]
\centering
\includegraphics[width=0.75\textwidth]{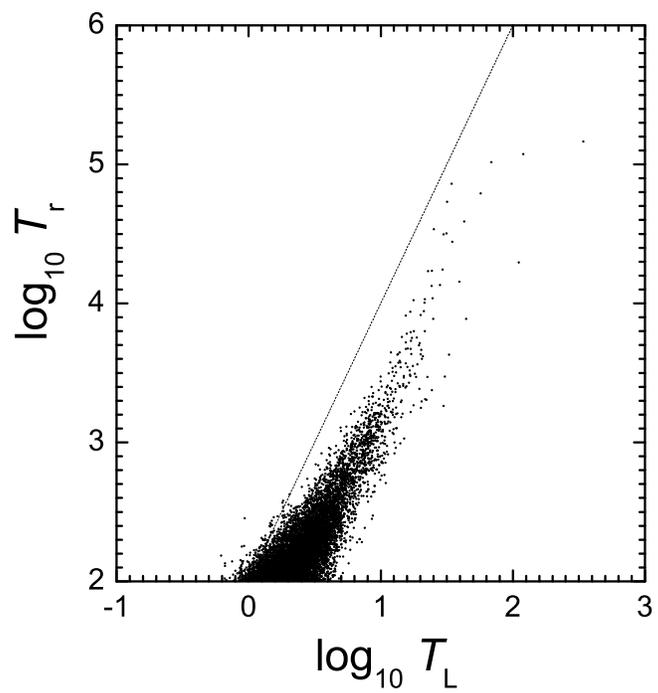}
\caption{A statistical relation
``$\log_{10} T_\mathrm{L}$~--- $\log_{10} T_\mathrm{r}$'',
where $T_\mathrm{L}$ and $T_\mathrm{r}$ are
measured in map time units; $\lambda = 5$. The straight dotted
line shows the quadratic dependence.}
\label{ttm}
\end{figure}

\begin{figure}[h!]
\centering
\includegraphics[width=0.75\textwidth]{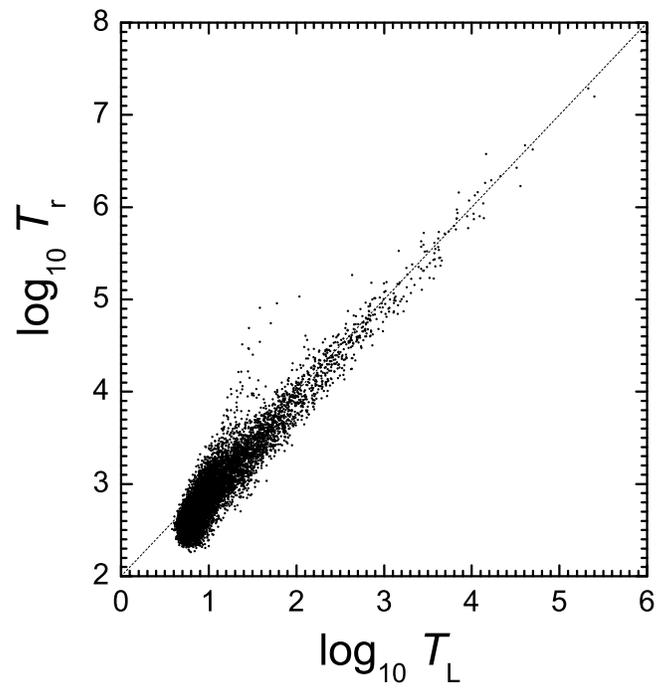}
\caption{The same as in Fig.~\protect{\ref{ttm}}, but the
recurrence times are measured in real time units. The straight
dotted line shows the linear dependence.}
\label{ttc}
\end{figure}

To illustrate our theoretical inferences, let us compute the
``$T_\mathrm{L}$~--- $T_\mathrm{r}$'' relationships for the Kepler
map in map time units (iterations) and, for comparison, in real
time units. The finite-time maximum Lyapunov exponent is computed
for a recurrence. In Fig.~\ref{ttm}, the computed relationship in
the map time units is shown for $\lambda = 5$, $n_{\rm it} =
10^{7}$. One can see, that, judging by the general slope of the
dependence in log-log scale, the dependence is far from being
linear here. Its slope in log-log scale is much steeper: the power
law index is equal to $1.5$--2, as expected. Now let us measure
the recurrence times in real time units. Apart from this change in
the time units, the same (as in Fig.~\ref{ttm}) dependence is
built in Fig.~\ref{ttc}. Now the relationship is evidently
quasilinear, in accord with our theoretical finding for the case
of real time units. One can even see how the ``spreading''
mechanism operates: the dependence has a ``{\it V}'' form, where
the left wing is much shorter and represents a remnant (left after
spreading by LF1) of the generic close-to-quadratic relationship.
Thus the general ``composite'' appearance of the
``$T_\mathrm{L}$~--- $T_\mathrm{r}$'' diagram (in real time units)
for the Kepler map mimics general structure of the
``$T_\mathrm{L}$~--- $T_\mathrm{r}$'' diagrams revealed in
computations of the disruption process in the three-body problem
in various settings; compare Fig.~\ref{ttc} with figures 2 and 3
in \cite{MT07} or with figures 3 and 7 in \cite{UH08}.

\begin{figure}[h!]
\centering
\includegraphics[width=0.75\textwidth]{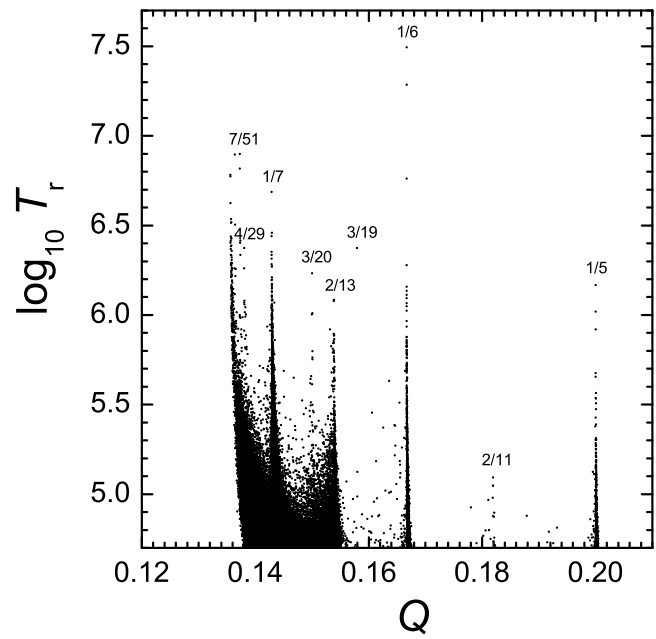}
\caption{A spectrum of winding numbers, visualizing the resonant
structure of the chaotic motion near chaos border; $\lambda = 5$.}
\label{swn}
\end{figure}

The over-all appearance of the diagram in Fig.~\ref{ttm} looks
rather irregular, in comparison with that in Fig.~\ref{ttc}. The
nature of this irregularity can be clarified by means of
construction of a spectrum of winding numbers \cite{S96}. It can
be built here for the Kepler map analogously to construction of
the spectrum for the ordinary separatrix map in \cite{S96}. The
spectrum of winding numbers graphically demonstrates which of the
resonant chains of islands produce the longest events of sticking.
The winding number $Q$ is formally defined for a recurrence as $Q
= \Delta x / n$, where $\Delta x$ is the length of a Poincar\'e
recurrence (the duration of an interval between crossings of the
separatrix), measured as the total sum of variations in the phase
$x$ {\it taken modulo $2 \pi$}, and $n = T_\mathrm{r}$ is the
length of the Poincar\'e recurrence measured in the map
iterations. We build the spectrum of winding numbers by plotting
$\log_{10} n$ versus $Q$. In Fig.~\ref{swn}, the spectrum of winding
numbers computed for $\lambda = 5$ and $n_{\rm it} = 5 \cdot
10^{11}$ is shown. The Farey tree \cite{M92} of resonances is
evident. Consider some lowest order resonances $m/n$ and $m'/n'$
that are ``neighboring'', i.e., $m n' - m' n = 1$, then the lower
level of the tree is made of ``mediants'' given by the formula
$m''/n''= (m + m')/(n + n')$. In Fig.~\ref{swn}, the evident
lowest order resonances are 1/7, 1/6 and 1/5. The mediants for
them are 2/13 and 2/11. For the resonances 1/7 and 2/13 the
mediant is 3/20; for the resonances 2/13 and 1/6 the mediant is
3/19; and so on. All mentioned resonances produce visible peaks
and are easily identified in Fig.~\ref{swn}. We see that the
irregular structure in Fig.~\ref{ttm} is explained by overlapping
of individual relationships for several sticky island chains. The
overlap of these relationships at the given timescale of
recurrences produces the observed irregularity.

\section{Range of applicability}
\label{rappl}

Let us consider in more detail the range of applicability of the
presented results on the statistics of the disruption process.
Clearly, these results are valid wherever the Kepler map
(\ref{km}) description of the motion of the tertiary is valid.
First of all, we assume that the orbit of the escaping body is
highly eccentric, and its pericenter distance $q$ is much greater
than the size of the orbit of the main binary.

Hence the inferred statistics of the disruption and Lyapunov times
are expected in an hierarchical restricted three-body problem,
where the pericenter distance of the tertiary is much greater than
the size of the orbit of the main binary. The eccentricity of the
orbit of the main binary as well as the non-coplanarity of the
three-body system do not play role due to the following reasons.
As it has been already mentioned above, the Kepler map is known to
describe the highly eccentric motion of a tertiary in several
different settings of an hierarchical restricted three-body
problem: in the external restricted planar and strongly non-planar
problems, also in the Sitnikov problem. If the pericenter distance
is large enough, only one harmonic in the Fourier expansion of the
energy increment is important and is enough to be taken into
account \cite{PB88,LS94}.

A role of the mass parameter $\mu$ in the main binary has not been
explored yet, but one may expect the validity of the Kepler map
approximation for the motion again, when $q$ is large, because the
higher order harmonics in the energy increment expansion are
exponentially small with $q$ \cite{PB88,LS94}. Moreover, the
formula for the energy increment in Eqs.~(\ref{km1}) in the case
of the circular-orbit main binary with the mass parameter $\mu =
1/2$ (equal-mass binary) is similar to that in the case of the
circular-orbit main binary with small $\mu$. To clarify this
point, let us compare the formulas. In the circular planar
restricted three-body problem with small $\mu$ one has for the
energy increment, if $q \gg 1$ ($q$ is measured in the units of
the semi-major axis of the main binary):
\begin{equation}
\Delta w \equiv w_{i+1} - w_i \propto \mu q^{-1/4} \exp \left( -
\frac{2^{3/2} q^{3/2}}{3} \right) \sin \tau_i
\label{dw1}
\end{equation}

\noindent in the case of prograde orbits of the tertiary, and
\begin{equation}
\Delta w \propto \mu q^{-7/4} \exp \left( - \frac{2^{3/2}
q^{3/2}}{3} \right) \sin \tau_i \label{dw2}
\end{equation}

\noindent in the case of retrograde orbits of the tertiary
\cite[formulas~(3.16)]{PB88}. In the case of the equal-mass main
binary and the non-planar problem, it can be found from
\cite[formula~(26)]{RH03} in the restricted problem limit that
\begin{equation}
\Delta w \propto q^{3/4} \exp \left( - \frac{2^{5/2} q^{3/2}}{3}
\right) \cos^4 \frac{I}{2} \sin 2 \tau_i , \label{dw3}
\end{equation}

\noindent where $I$ is the inclination of tertiary's orbit. We see
that the structure of the formulas is similar, putting aside some
differences in the numerical values of the coefficients and the
power-law indices. Note that formulas~(\ref{dw1}, \ref{dw2},
\ref{dw3}) can be used, if necessary, to find numerical values of
the $\lambda$ parameter in Eqs.~(\ref{km}), because these formulas
provide estimates of the $W$ parameter in Eqs.~(\ref{km1}), while
the formula for $\kappa$ in Eqs.~(\ref{km1}) is trivial (it is
simply the time normalization, see, e.g., \cite{PB88}).

The inferred statistics appear to be still valid in a more general
``hierarchical resonant scattering'' \cite{H75,H93} setting for a
three-body interaction, where the masses of ``stars'' are
arbitrary. Hut \cite{H93} derived an heuristic
$T_\mathrm{d}^{-2/3}$
law as a lower bound for the time decay of the survival
probability in
this setting, and showed this law to describe well the tails of
numerical-expe\-rimental distributions. Note that this
heavy-tailed distribution is in accord with an early finding by
Agekian et al.\ \cite{A83} that the mean life-time of a general
isolated three-body system is infinite.

As we have seen above, two basic assumptions are necessary
for heuristic derivation of the $T_\mathrm{d}^{-2/3}$ law,
namely, (i)~that the distribution of ejection energies is smooth
in the neighborhood of the energy threshold $E=0$, (ii)~that the
asymptotic decay of the survival probability is the same as the tail
of the distribution of the orbital periods of the escaping body.
These two assumptions look rather plausible even in the general
three-body problem. Combining these considerations and our
theoretical findings described above, we see that the
$T_\mathrm{d}^{-2/3}$ law is expected to be quite universal.

However, neither the $T_\mathrm{d}^{-2/3}$ law, nor even any other
algebraic law were reported in the numerical-expe\-rimental
studies in \cite{MT07,UH08}. We think that the main reasons are
that the algebraic fitting functions were not used, and, besides,
solely the initial part of the distribution was built in
\cite{MT07}, though the time range of the simulations allowed one
to study the tails. Another point is that the tail of the
disruption time distribution in the given problem should be
considered separately from the initial part, because it
corresponds to a different dynamical situation: here the regime of
decay might be Poissonian (see analogues in \cite{SS96,SS97,S99}),
or, in the very beginning, inverse square-root
\cite{CS81,C99,S99}.

Our theoretical inferences seem to be confirmed by the
results of the very recent study \cite{ORS10}, where the
statistics of the decay process in the equal-mass three-body
problem with randomized initial conditions were investigated in
extensive numerical experiments. The lifetime distributions
obtained in \cite{ORS10} have turned out to be heavy-tailed, i.e.,
the tails have turned out to be algebraic. The computed power-law
index $\alpha$ for the integral distribution has been found to be
within the narrow range, approximately from $0.4$ to $0.7$,
depending on the virial coefficient (see \cite{ORS10}). The
theoretically predicted value $\alpha = 2/3$ is within this narrow
range.

The range of applicability of the derived ``$T_\mathrm{L}$~---
$T_\mathrm{d}$'' relationship is similar to that of the derived
distribution law. The proximity of our theoretical quasilinear
``$T_\mathrm{L}$~--- $T_\mathrm{d}$'' relationship to the
numerical-expe\-rimental results in \cite{MT07} (obtained for an
equal-mass three-body system) has a natural heuristic explanation,
consisting in that the stage before disruption is hierarchical,
with the outer body exhibiting the usual final L\'evy flights, and
this process can be described in the ``hierarchical resonant
scattering'' setting for a three-body interaction (though, in
general, triple encounters can play role).

Finally, there is no wonder that a simple one-parameter
two-dimensional map, such as the Kepler map, is able to describe
the essential dynamics of disruption of a system with several
degrees of freedom. The matter is that we consider a very special
stage of the system evolution subject to serious limitations: the
orbit of the escaping body is highly eccentric, and its pericenter
distance is much greater than the size of the main binary. In many
respects, reduction of the motion to the Kepler map in the problem
considered is similar to reduction of the motion in the vicinity
of the separatrices of the ``guiding resonance'' to the separatrix
map (see \cite{C79}) in a general Hamiltonian system. An important
difference, however, is that the ordinary separatrix map has two
parameters, and cannot be reduced, opposite to the case of the
Kepler map, to a one-parameter form, because the phase increment
in the ordinary separatrix map is logarithmic.

\section{Conclusions}

We have considered statistics of the disruption and Lyapunov times
in an hierarchical restricted three-body problem. As we have seen,
at the edge of escape the orbital periods of the escaping body
exhibit L\'evy flights. Due to them, the distribution of the
disruption times is heavy-tailed with the power-law index of the
integral distribution equal to $-2/3$ , while the relation between
the Lyapunov and disruption times is quasilinear. The former
finding is in accord with heuristic and numerical-expe\-rimental
results in \cite{SB01,H93,MT99}, while the latter one is in
accord with recent numerical-expe\-rimental results in
\cite{MT07,UH08}. Our theoretical results are valid for any system
described by the Kepler map, i.e., they are valid at least in the
external restricted three-body problem, both planar and
non-planar, where the tertiary does not suffer close encounters
with the central binary. The derived statistical laws appear to be
valid as well in a more general ``hierarchical resonant
scattering'' setting for a three-body interaction.

The sharp difference between the two kinds of Hamiltonian
intermittency, in what concerns the slope indices of the
asymptotic power-law distributions of the Poincar\'e recurrences,
allows one to explain the observed difference in the power-law
indices of the distribution laws reported for the chaotic dynamics
of the Solar system minor bodies. Dones et al.\ \cite{DLD96}
reported on the algebraic tails of the integral distributions with
the power-law index $\alpha$ equal to $0.8 \pm 0.2$, whereas
Shevchenko and Scholl \cite{SS96,SS97} reported on the tails with
the index equal to $\approx 1.5$. In \cite{DLD96} the escape times
in the highly-eccentric chaotic cometary dynamics in the Solar
system were studied (the perturbations due to the four giant
planets were taken into account), whereas in \cite{SS96,SS97} the
subject was the low-eccentricity intervals between the
eccentricity jumps in the chaotic asteroidal dynamics (in the
restricted three-body problem Sun--Jupiter--asteroid). Judging by
the values of the power-law index, the former statistics
correspond to the Hamiltonian intermittency of the first kind,
whereas the latter one to that of the second kind. The statistics
are LF1-dominated and LF2-dominated, respectively, and the predicted
power-law indices $\alpha$ are respectively equal to $2/3$ and
$\approx 3/2$.
The evident ``inverse symmetry'' of the indices is a property of
the gravitational dynamics, described by the Kepler map;
generally, when a separatrix map with an arbitrary $\gamma$ is in
action, the predicted indices are $1/\gamma$ and $\approx 3/2$,
respectively. Thus the latter index is more ``universal'' than the
former one.

The change of the power-law index value ($\approx 1$ instead of
$\approx 2$) in the ``$T_\mathrm{L}$~--- $T_\mathrm{d}$''
relationship in comparison with the results on the dynamics of the
Solar system minor bodies obtained
in~\cite{SFL90,LFM92,MLF94,LD93,FM97,S98a,TVD05} is due to the
fact that the singularity at crossing the separatrix is much
stronger in the considered problem (which is described by the
Kepler map instead of the ordinary separatrix map), and therefore
the first kind Hamiltonian intermittency dominates over its second
kind and thus defines the properties of the ``$T_\mathrm{L}$~---
$T_\mathrm{d}$'' relationship and the tail of the disruption time
distribution.

We have shown that a critical non-zero value
$\gamma_\mathrm{crit}$ of the $\gamma$ parameter of general
map~(\ref{km}) exists that separates the maps with LF2-dominated
dynamics from those with LF1-dominated dynamics.

As a by-product of our study we have proposed a new method for
estimating the critical exponent $\alpha$ in Eq.~(\ref{fta}). This
method is based on computation of the $\gamma$ transition value
separating maps~(\ref{km}) with LF1-dominated and LF2-dominated
statistics of long recurrences.

Finally, our inferences shed light on the mechanism of disruption
of an hierarchical three-body system in the adopted setting of the
problem: the typical way of disruption, as described by the Kepler
map, is a kind of ``L\'evy unfolding'' of the system in both time
and space: at the edge of the system's disruption, the escaping
body exhibits L\'evy flights in its orbital period and semi-major
axis, and in the course of this random process the orbital period
and semi-major axis become arbitrarily large until the separatrix
separating the bound and unbound states of the motion is crossed
and the body escapes.

\section*{Acknowledgments}

The author is thankful to V.~V.~Orlov and anonymous referees for
valuable remarks and comments. This
work was partially supported by the Russian Foundation for Basic
Research (projects \# 09-02-00267 and \# 10-02-00383) and by the
Programme of Fundamental Research of the Russian Academy of
Sciences ``Fundamental Problems in Nonlinear Dynamics''.

\newpage

\end{document}